\documentclass[prd,twocolumn,reprint,preprintnumbers,nofootinbib,superscriptaddress]{revtex4-1}
\usepackage{eqnarray}
\usepackage{soul}

\input{header}

\begin{document}

\title{Machine Learning the Higgs-Top CP Phase}

\preprint{{\footnotesize DESY 21-161}}

\author{Rahool Kumar Barman}\email{rahool.barman@okstate.edu}
\affiliation{Department of Physics, Oklahoma State University, Stillwater, OK, 74078, USA}
\author{Dorival Gon\c{c}alves}\email{dorival@okstate.edu}
\affiliation{Department of Physics, Oklahoma State University, Stillwater, OK, 74078, USA}
\author{Felix Kling}\email{felixk@slac.stanford.edu}
\affiliation{SLAC National Accelerator Laboratory, 2575 Sand Hill Road, Menlo Park, CA 94025, USA}
\affiliation{Deutsches Elektronen-Synchrotron DESY, Notkestrasse 85, 22607 Hamburg, Germany}

\begin{abstract}
We explore the direct Higgs-top CP  measurement via the $pp\to t\bar{t}h$ channel at the high-luminosity LHC. We show that a combination of machine learning techniques and efficient kinematic reconstruction methods can boost new physics sensitivity, effectively probing the complex $t\bar{t}h$ multi-particle phase space. Special attention is devoted to top quark polarization observables, uplifting the analysis from a raw rate to a polarization study. Through a combination of hadronic, semi-leptonic, and di-leptonic top pair final states in association with $h\to \gamma\gamma$, we obtain that the HL-LHC can probe the Higgs-top coupling modifier and CP-phase, respectively,  up to $|\kappa_t|\lesssim 8\%$ and $|\alpha|\lesssim 13^{\circ}$ at $68\%$~CL.
\end{abstract}

\maketitle

\section{Introduction}
\label{sec:intro}

New sources of CP violation can be a key ingredient to explain the matter-antimatter asymmetry of the universe~\cite{Sakharov:1967dj,Kajantie:1996mn,Huet:1994jb}. Hence, the quest for new CP violating interactions is a clear target for beyond the Standard Model (SM) searches, being a critical component of the physics program of the LHC. A particularly interesting option is that the Higgs boson couplings present these new physics sources. 

From the theoretical point of view, some Higgs interactions are more inclined to display CP violation effects than others. While the widely studied beyond the SM CP structure for the Higgs to vector boson couplings are loop suppressed, arising only at dimension-6 or higher~\cite{Buchmuller:1985jz,Grzadkowski:2010es}, CP violation in Higgs to fermion interactions can manifest already at the tree-level~\cite{Buckley:2015vsa}, being naturally larger. Owning to its magnitude, the top quark Yukawa coupling can play a significant role in this context and be most sensitive to new physics.

Whereas it is possible to access the Higgs-top coupling through loop induced processes~\cite{Brod:2013cka,Dolan:2014upa,Englert:2012xt,Kobakhidze:2016mfx,Bernlochner:2018opw,Englert:2019xhk,Gritsan:2020pib,Bahl:2020wee}, the direct Higgs-top production via $pp\to t\bar{t}h$ is crucial to disentangle possible new physics effects~\cite{Ellis:2013yxa,Boudjema:2015nda,Buckley:2015vsa,Buckley:2015ctj,Gritsan:2016hjl,Goncalves:2016qhh,Mileo:2016mxg,AmorDosSantos:2017ayi,Azevedo:2017qiz,Li:2017dyz,Goncalves:2018agy,ATLAS:2018mme,CMS:2018uxb,Ren:2019xhp,Bortolato:2020zcg,Cao:2020hhb,MammenAbraham:2021ssc,Martini:2021uey,Goncalves:2021dcu}. This channel was observed in 2018 by both ATLAS and CMS with significances of 6.3~$\sigma$ and 5.2~$\sigma$, respectively~\cite{ATLAS:2018mme,CMS:2018uxb}. The high-luminosity LHC (HL-LHC) studies indicate that the Higgs-top interaction will be probed to outstanding accuracy at the end of the LHC run, reaching $\delta y_t\lesssim 4\%$ when combining the HL-LHC ATLAS and CMS data~\cite{Cepeda:2019klc}. The same projections indicate that the $t\bar{t}h$ channel in the $h\to \gamma\gamma$ final state will display dominant sensitivities. While the di-photon final state presents limited statistics, it highly benefits from controlled backgrounds from side-bands.  Recently, ATLAS and CMS have reported the first experimental Higgs-top CP studies, exploring the $t\bar{t}h$ channel~\cite{ATLAS:2020ior,CMS:2020cga}. Both analyses focus  on the di-photon final state, $h\to \gamma\gamma$. ATLAS and CMS exclude Higgs-top CP-mixing angles above $43^\circ$ and $55^\circ$ at 95\%~CL, respectively. 

In the present manuscript, we perform a detailed investigation of the Higgs-top CP sensitivity with the $pp\to t\bar{t}h$ channel at the HL-LHC, considering the most promising decay mode, $h\to \gamma\gamma$. We explore the complex multiparticle final state with a combination of machine learning techniques and efficient kinematic reconstruction methods. Since distinct Higgs-top CP-phases affect the net top and anti-top quark polarization, propagating the spin effects to the  top quark final states, we devote special attention to include the top polarization observables in our study. In particular, these spin effects are used to define genuine CP-observables. After motivating and constructing the relevant kinematic observables, we evaluate how much information can be extracted with them. The convenient metric adopted to quantify this is given by the Fisher information. We show that the ability of probing the $pp\to t\bar{t}h$ channel exploring the complex multiparticle final state  not only in terms of a raw rate, but as a polarized process, can offer a crucial pathway to probe the underlying production dynamics, accessing possible new physics effects.

The structure of this paper is as follows. In Section~\ref{sec:theory}, we present the theoretical parametrization for the top Yukawa coupling. We discuss the new physics effects to the top polarization, define the CP-sensitive observables, and quantify how much information on the CP-phase can be extracted using distinct observables.  In Section~\ref{sec:kin_reconstruct}, we present the kinematic reconstruction methods, which are relevant to build  prominent observables to new physics and maximally explore the $t\bar{t}h$ final state. Next, in Section~\ref{sec:analysis}, we move on to the detailed analysis, where we derive the projected sensitivities for the Higgs-top CP-phase at the HL-LHC. This study is inclusive in respect to the top pair final states, combining the leptonic, semi-leptonic, and hadronic channels.  Finally, a summary of our key findings is delivered in Section~\ref{sec:summary}.

\section{CP Structure and Observables}
\label{sec:theory}

We parametrize the top quark Yukawa coupling with the following Lagrangian
\be
    \mathcal{L} = - \frac{m_{t}}{v} \kappa_t \bar{t}(\cos\alpha + i \gamma_5 \sin\alpha) t h\, ,
    \label{eq:Higgs-top}
\ee
where $m_{t}$ is the mass of the top quark, $v$ is the vacuum expectation value in the SM~($v=246$~GeV), $\kappa_{t}$ is a real number, and $\alpha$ is the CP-phase. The interaction between the CP-even Higgs boson and the top quark in the SM is represented by $(\kappa_{t},\alpha)=(1,0)$, while $\alpha=\pi/2$ results in a pure CP-odd Higgs-top interaction. New physics contributions in \cref{eq:Higgs-top} will display effects both in the Higgs $t\bar{t}h$ production  and  decay, $h\rightarrow \gamma\gamma$. Whereas  the Higgs decay will more relevantly change the total signal rate~\cite{Ellis:2013yxa},  we will devote special attention to probe the new physics effects in the Higgs production, exploring the top quarks' final state kinematics. This will be an essential ingredient to uplift the new physics sensitivity from  CP-phase effects.

\begin{figure*}[!htb]
    \centering
   \includegraphics[height=4.55cm,width=5.8cm]{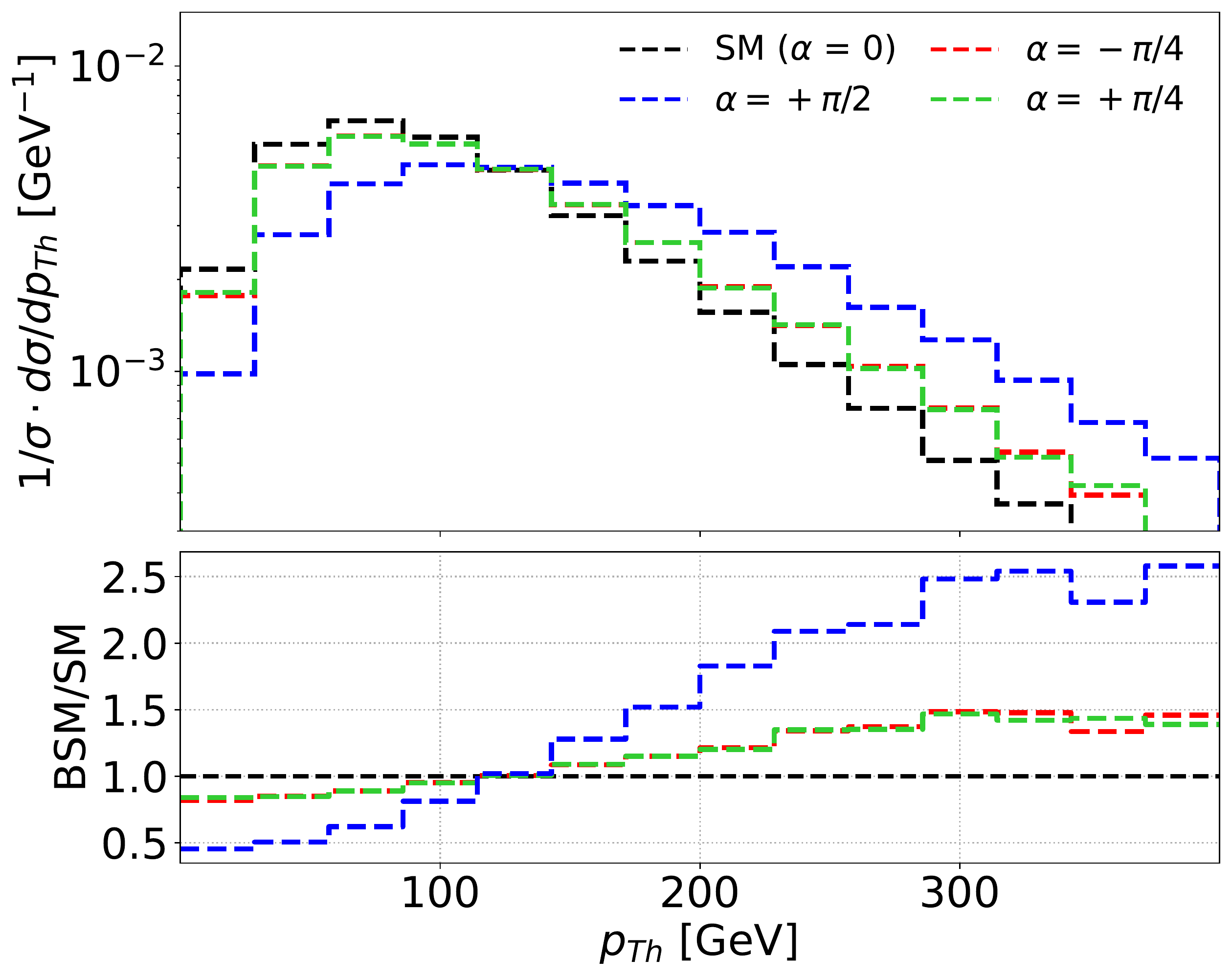} 
   \includegraphics[height=4.55cm,width=5.8cm]{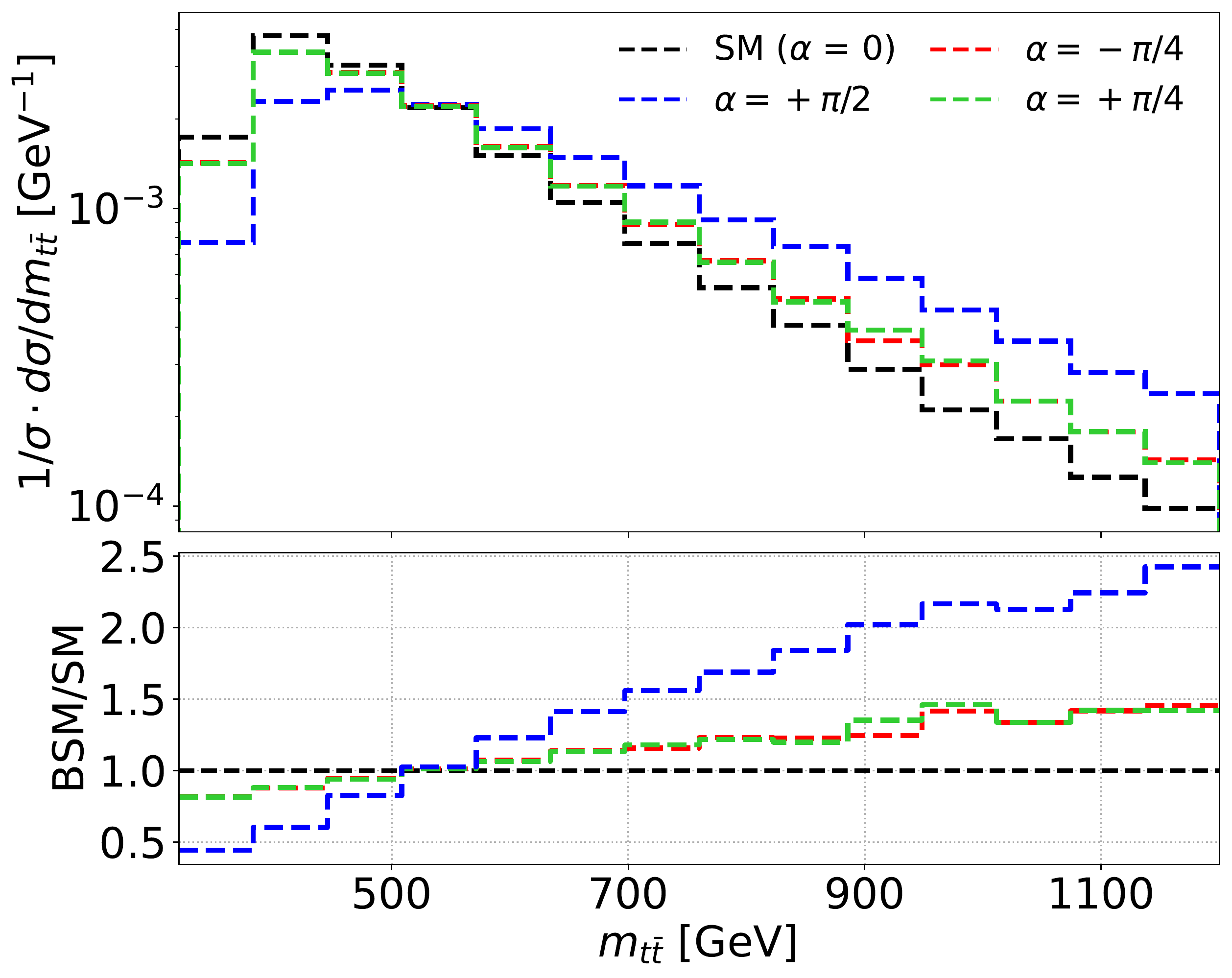} 
   \includegraphics[height=4.55cm,width=5.8cm]{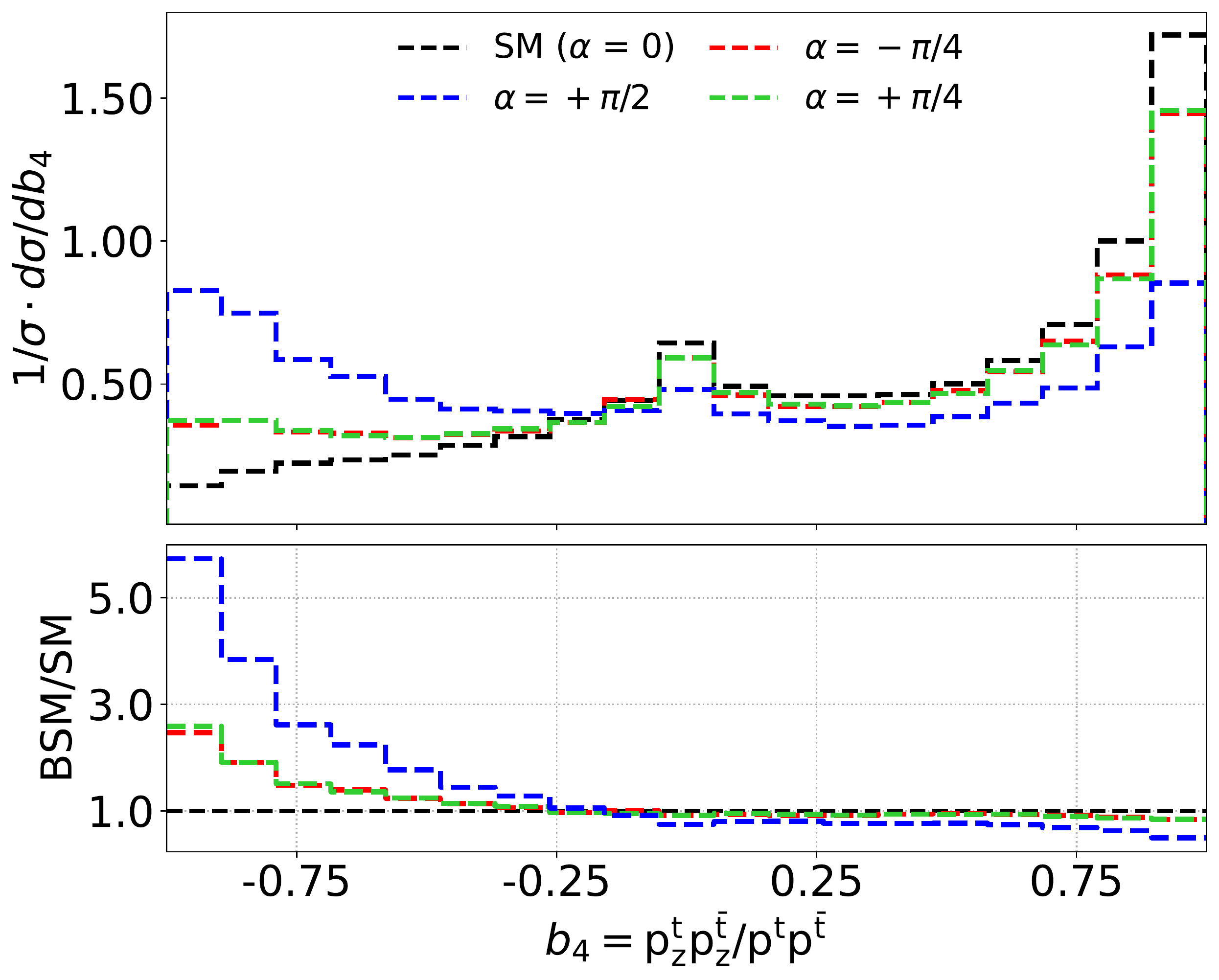}  \\ 
   \includegraphics[height=4.55cm,width=5.8cm]{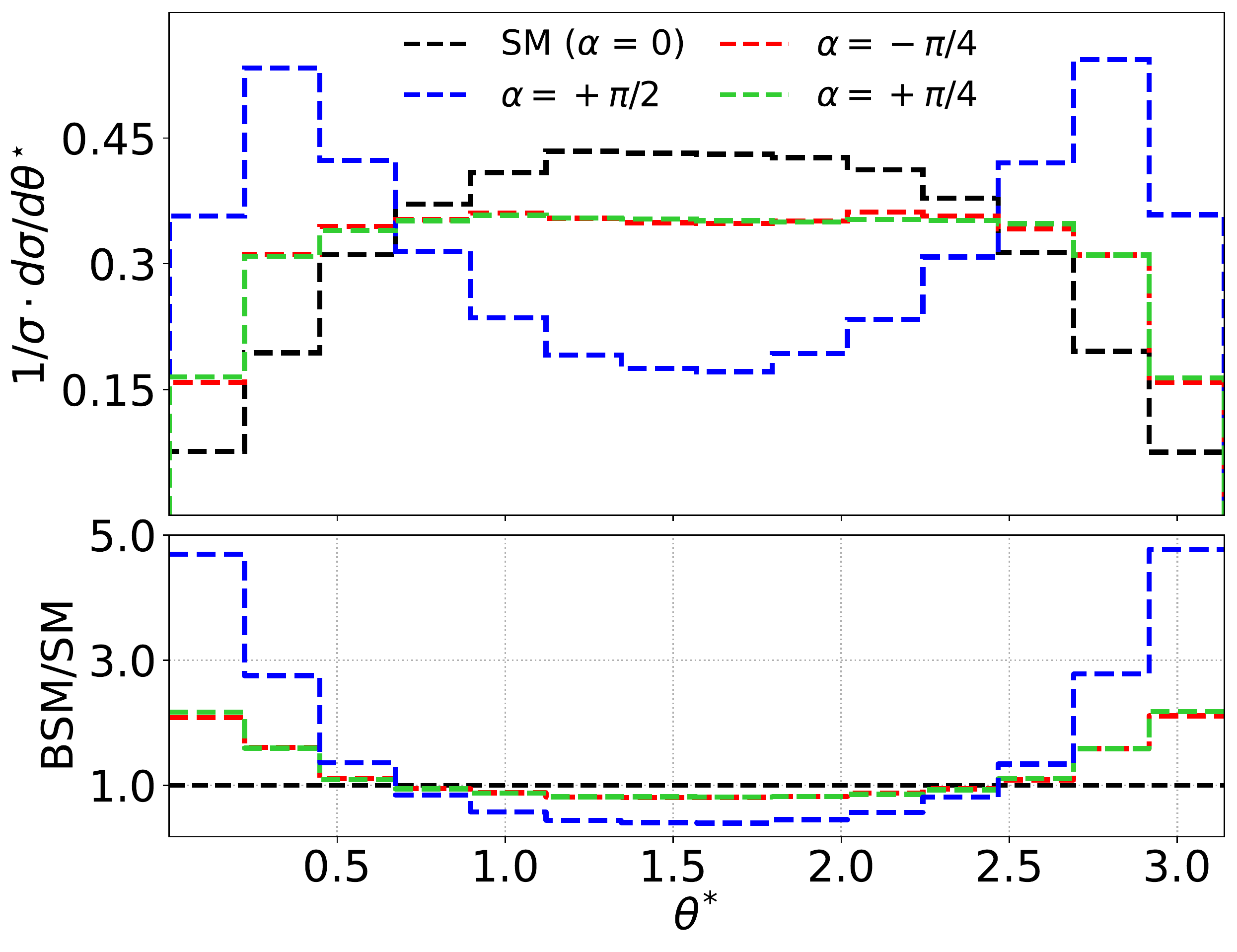}    
   \includegraphics[height=4.55cm,width=5.8cm]{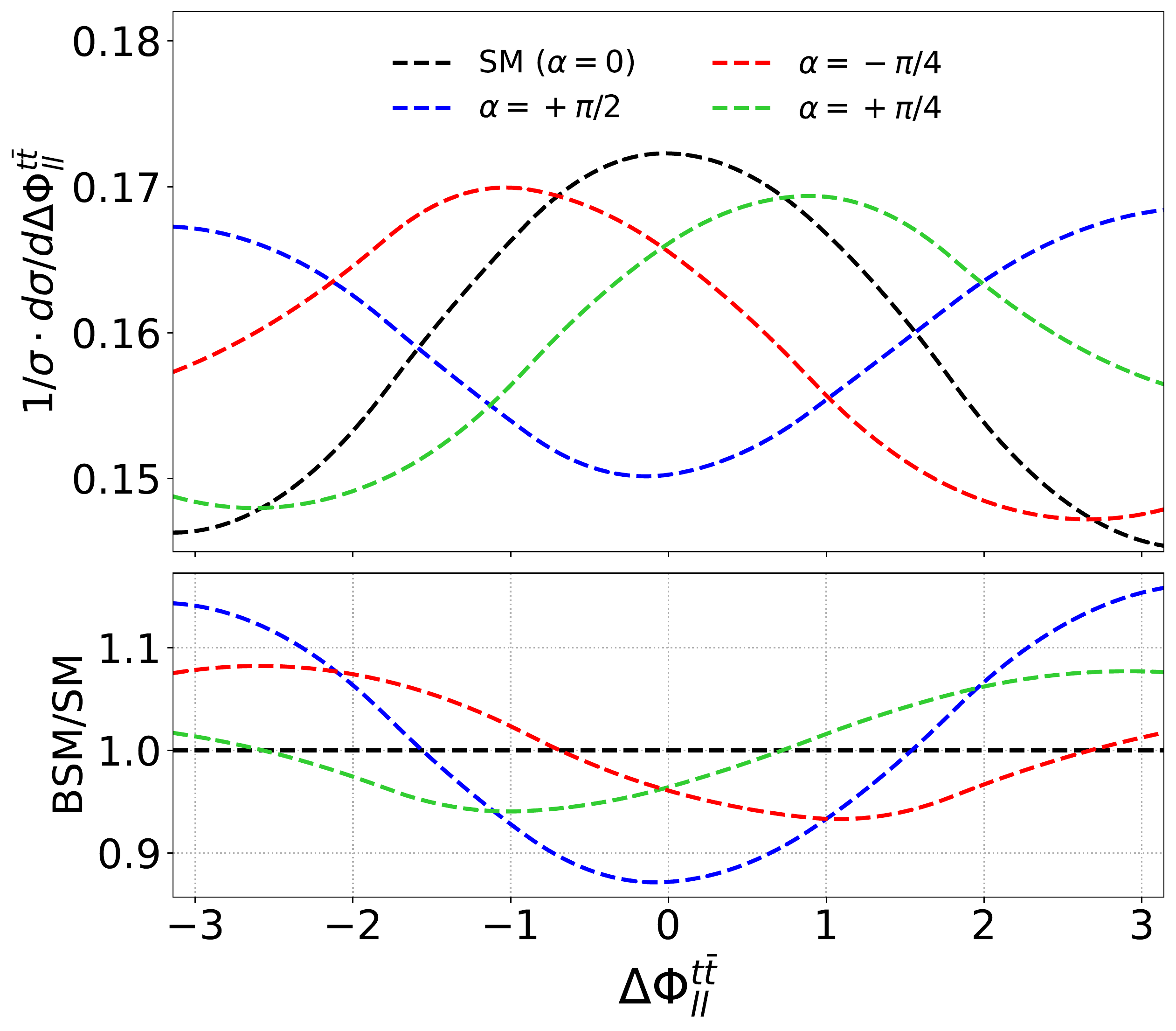} 
   \includegraphics[height=4.55cm,width=5.8cm]{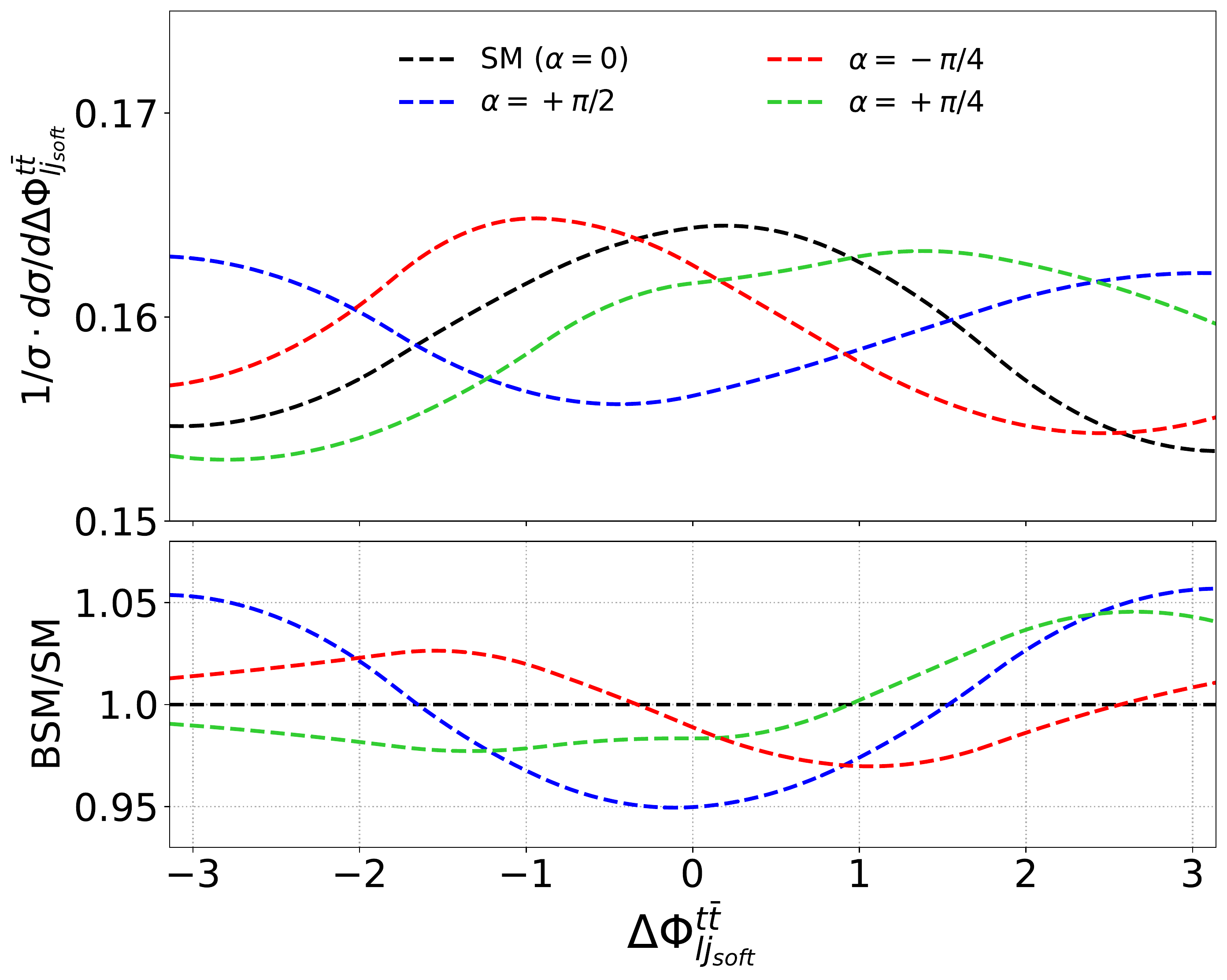}
    \caption{Top panels: Distributions for the transverse momentum for the Higgs boson $p_{Th}$ (left), invariant mass for the top pair $m_{t\bar{t}}$ (center), and the product of the projections of the top and anti-top momentum $b_{4} = p_{t}^{z} p_{\bar{t}}^{z} / p_{t} p_{\bar{t}}$ (right). Bottom panels: Distributions for the Collins-Soper angle $\theta^{*}$ (left), the azimuthal angle between the two charged leptons in the top pair rest frame $\Delta\phi_{\ell \ell}^{t\bar{t}}$ for fully leptonic $t\bar{t}h$ events (center), and the same angle between the charged lepton and the softest light jet in the top rest frame $\Delta\phi_{\ell j_\text{soft}}^{t\bar{t}}$ for semi-leptonic $t\bar{t}h$ events (right).  Each panel shows parton level results for the $t\bar{t}h$ sample for the SM Higgs ($\alpha = 0$), a CP-odd Higgs ($\alpha = \pi/2$) and mixed hypotheses ($\alpha = \pm\pi/4$). We also present the ratio between new physics and SM scenarios on the bottom panel of each figure. The results are presented for the 14~TeV LHC.}
    \label{fig:semilep_parton_dist}
\end{figure*}

\subsection{Top Polarization}

Among the observables sensitive to the structure of the top quark Yukawa interaction, the spin correlations between the top and anti-top in $t\bar{t}h$ associated production offer a prominent pathway for precision studies~\cite{Bar-Shalom:1995quw,Gunion:1996xu, Atwood:2000tu,Berge:2008wi, Ellis:2013yxa,  Buckley:2015vsa, Boudjema:2015nda, Gritsan:2016hjl, Buckley:2015ctj,Goncalves:2016qhh,Mileo:2016mxg, Goncalves:2018agy, Aad:2020ivc,Goncalves:2021dcu,Degrande:2021zpv}. Owing to its short lifetime ($\sim 10^{-25}$s)~\cite{ParticleDataGroup:2020ssz}, the top quark is expected to decay before hadronization occurs ($\sim 10^{-24}$s) and spin decorrelation effects take place ($\sim 10^{-21}$s)~\cite{Mahlon:2010gw}. Thus, the spin-spin correlations between $t$ and $\bar{t}$ can be traced back from the top quark decay products. In particular, it is possible to observe correlations between any two decay products, one from the top quark and the other from the anti-top quark. The correlations scale with the spin analyzing power associated with each top decay product. 

More accurately, the top quark final states in the leptonic $t\to W^+ b\to \ell^+ \nu b$ and hadronic $t\to W^+ b\to \bar{d} u b$ channels are correlated with the top quark spin axis as follows: 
\be
    \frac{1}{\Gamma}\frac{d\Gamma}{d\cos\xi_i}=\frac{1}{2}\left( 1+ \beta_i P_t\cos\xi_i \right)\,,
\ee
where $\Gamma$ is the partial decay width, $\xi_i$ is the angle between the $i$-th decay product and the top quark spin axis in the top quark rest frame, $P_t$ is the polarization of the decaying top $(-1\le P_t\le 1)$, and $\beta_i$ is the spin analyzing power of the final state particle $i$~\cite{Bernreuther:2010ny}. At leading order, the coefficient $\beta_i$ is $+1$ for charged lepton $\ell^+$ and $\bar{d}$-quark, $-0.3$ for $\bar{\nu}$ and $u$-quark, $-0.4$ for the $b$-quark, and $0.4$ for $W$-boson. The sign of the coefficient $\beta_i$ is flipped for anti-top decays.

Granted by the $V\!-\!A$ current structure of the weak interaction, the charged lepton will be a prominent spin analyzer, favoring studies with di-leptonic top pairs. Exploring this phenomenology, the $\Delta \phi_{\ell\ell}^\text{lab}$ observable, which is the azimuthal angle difference between the two charged leptons in the lab frame,  is a good example of probe that has  been found effective in accessing the Higgs-top CP-properties~\cite{Buckley:2015vsa,Goncalves:2018agy}. Remarkably, the sensitivity of $\Delta \phi_{\ell\ell}^\text{lab}$ improves further in the boosted Higgs regime due to the change in the net polarization for the top-pair at high energies. 

Analogously to the charged lepton, the $d$-quark also presents maximal spin analyzing power. However, it is a challenging task to tag a $d$-quark jet in a collider environment.  An efficient solution is to select the softest of the two light-quark jets, $j_\text{soft}$,  in the top quark rest frame. This choice uplifts the spin analyzing power of  $j_\text{soft}$ to 50\% of the lepton's~\cite{Jezabek:1994qs}. This approach boosts the spin correlation analyses for the semi-leptonic and hadronic top quark pairs. Several  observables can be defined exploiting this fact, a particularly relevant example, that we will explore in this manuscript, is the azimuthal angle difference between the charged lepton and softest light jet in the top pair rest frame,~$\Delta\phi_{\ell j_\text{soft}}^{t\bar{t}}$. 

\subsection{CP-sensitive Observables}

Various  kinematic observables have been studied in the literature to access the Higgs-top CP structure in the $pp\to t\bar{t}h$ channel. Some illustrative distributions are presented in \cref{fig:semilep_parton_dist}, such as the transverse momentum for the Higgs boson $p_{Th}$ (top left)~\cite{Demartin:2014fia,Demartin:2015uha}, the invariant mass for the top pair $m_{tt}$ (top center), the product of projections of top and anti-top momentum $b_{4} = p_{t}^{z} p_{\bar{t}}^{z} / p_{t}p_{\bar{t}}$ (top right)~\cite{Gunion:1996xu}, and the angle between the top quark and the beam direction in the $t\bar{t}$ CM frame $\theta^{*}$ which is also known as Collins-Soper angle (bottom left)~\cite{Goncalves:2018agy}. These observables result in distinct profiles for different Higgs-top CP-phases. The pure CP-odd phase, $\alpha=\pm\pi/2$, leads to a shift to higher energies in the peak of the distributions compared to the SM scenario, $\alpha=0$. Different CP-phases interpolate between these two profiles without sensitivity for the sign of the phase.

The variables  $p_{Th}$, $m_{t\bar{t}}$, $b_{4}$, and $\theta^{*}$ are CP-even observables, being sensitive to the squared terms: $\cos^{2}\alpha$ and $\sin^{2}\alpha$. Thus, these probes are indifferent to the CP-even and CP-odd Higgs-top interference terms, which are proportional to $\cos\alpha \sin\alpha$. In particular, they are not sensitive to variations from a relative sign-difference in the CP-phase. Genuine CP-sensitive observables can be constructed from antisymmetric tensor products that require four linearly independent four-momenta. Owing to the top polarization being carried out to the decays, it is possible to construct such observable using, for instance, the top, anti-top and their decay products~\cite{Boudjema:2015nda, Mileo:2016mxg, Goncalves:2018agy}. In general, the antisymmetric tensor product can be expressed as
\be
    \epsilon(p_t,p_{\bar{t}},p_{i},p_{k}) \equiv \epsilon_{\mu\nu\rho\sigma}p_{t}^{\mu}p_{\bar{t}}^{\nu}p_{i}^{\rho}p_{k}^{\sigma}\,,
    \label{eq:eps_tensor_prod}
\ee
where $\epsilon_{0123} = 1$, and  $\{i,k\}$ represent the final state particles produced from the top and the anti-top decays, respectively.

In the $t\bar{t}$ CM frame, \cref{eq:eps_tensor_prod} can be fortuitously  simplified to $p_{t}\cdot(p_{i} \times p_{k})$. This mathematical relation can be used to define azimuthal angle differences between the decay products, in the $t\bar{t}$ CM frame, that are odd under CP-transformations~\cite{Goncalves:2018agy}:
\be
    \!\!\! \Delta \phi_{ik}^{t\bar{t}} \!=\!
    \text{sgn} \left[\vec{p}_{t} \!\cdot\! (\vec{p}_{i} \!\times\! \vec{p}_{k})\right] 
   \arccos \!\left[ \frac{\vec{p}_{t} \!\times\! \vec{p}_{i}}{|\vec{p}_{t} \!\times\! \vec{p}_{i}|} \!\cdot\! \frac{\vec{p}_{t} \!\times\! \vec{p}_{k}}{|\vec{p}_{t} \!\times\! \vec{p}_{k}|}\right]\!.
    \label{eq:CP_odd_obseravable}
\ee
For illustration, we present in \cref{fig:semilep_parton_dist} the azimuthal angle between the two charged leptons $\Delta\phi_{\ell \ell}^{t\bar{t}}$ in the fully leptonic case (bottom center) and between the charged lepton and the softest light jet in the top rest frame $\Delta\phi_{\ell j_\text{soft}}^{t\bar{t}}$ in the semi-leptonic case (bottom right). Two comments are in order. First, we notice that $\Delta \phi_{ik}^{t\bar{t}}$ is indeed sensitive to the sign of the CP-phase, as illustrated in a comparison between the distribution profiles for $\alpha=\pi/4$ against $-\pi/4$. Second, in light of the spin analyzing power of the charged lepton in relation to $j_\text{soft}$, the relative CP-sensitivity of the di-leptonic against the semi-leptonic correlation follows our expectation. Namely, the beyond the SM effects in the $\Delta\phi_{\ell j_{soft}}^{t\bar t}$ observable are $\sim 50\%$ weaker in respect to $\Delta\phi_{\ell\ell}^{t\bar t}$. This can be observed by comparing  the bottom panel of these plots, where we display the BSM/SM ratio.

\subsection{Observable Information}

\begin{figure*}[!htb]
   \centering
   \includegraphics[width=0.48\textwidth]{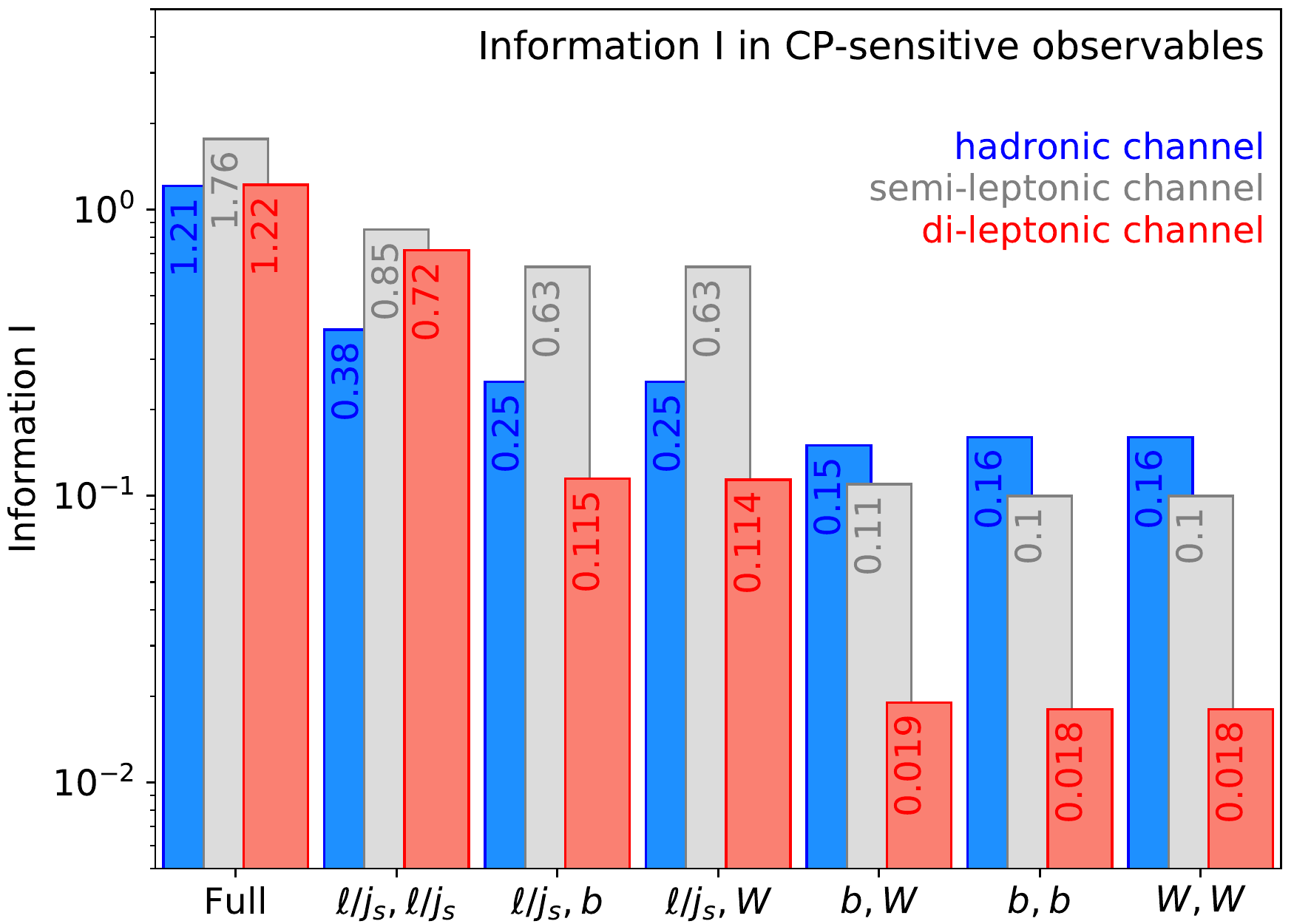}
   \includegraphics[width=0.48\textwidth]{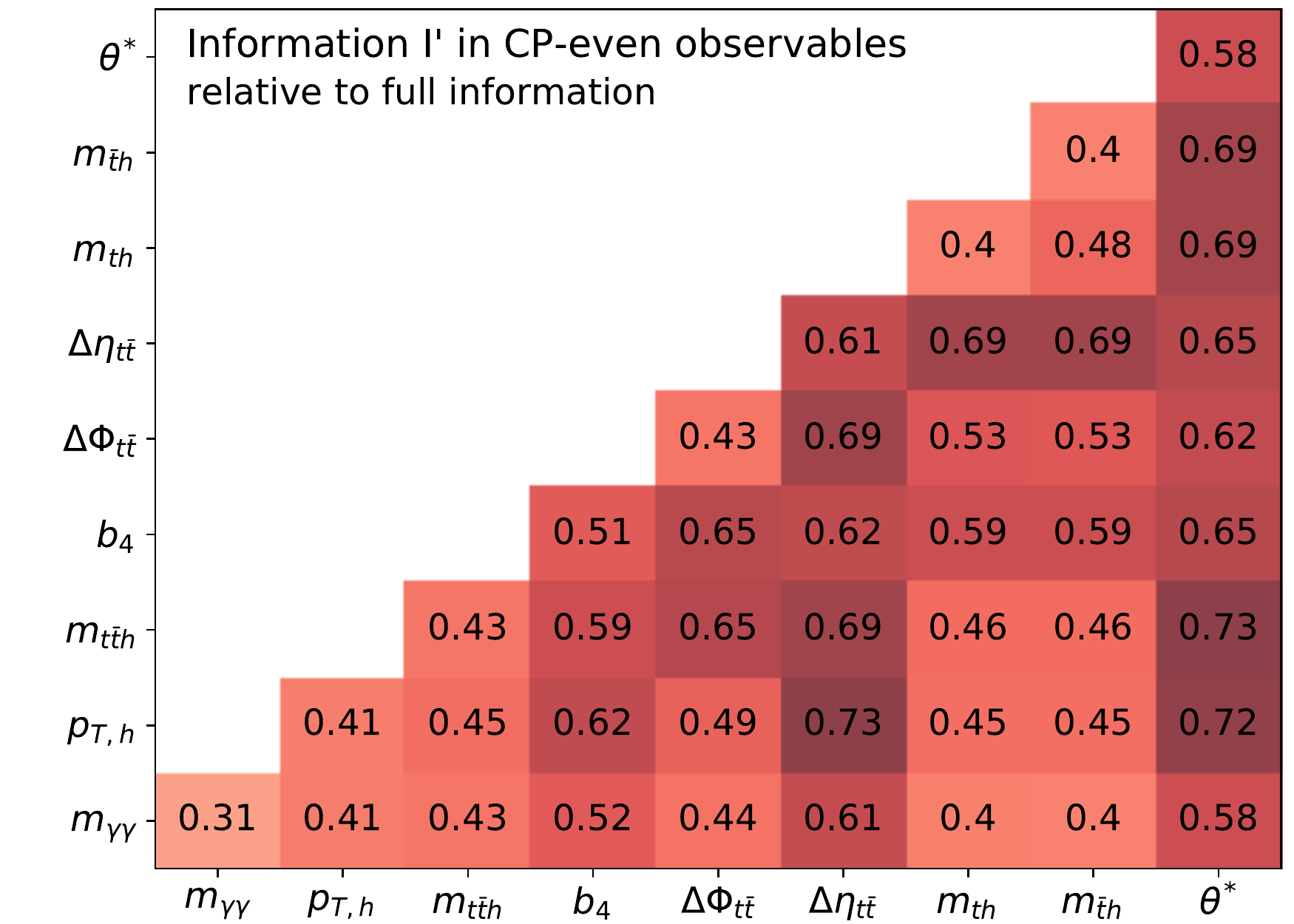}
    \caption{Comparison of sensitivity on the CP-phase $\alpha$ arising from different observables in terms of the Fisher Information $I$ for CP-odd observables probing linear new physics effects (left) and in terms of the modified Fisher Information $I'$ for CP-even observables probing the non-linear new physics effects (right).}
    \label{fig:information}
\end{figure*}

Before proceeding to a full analysis, let us pause for a moment to better understand which distributions and channels are sensitive to the CP-phase $\alpha$. In particular, we would like to quantify and compare how much information on the CP-phase is available using the different observables in a parton level setup. This will provide some benchmarks and highlight the main ingredients required for an efficient analysis strategy that will be presented in Sec.~\ref{sec:analysis}.

Let us first consider the spin correlation observables $\Delta \phi_{ik}^{t\bar t}$ between two decay products from the top and anti-top, which probe the new physics effects linear in $\alpha$. A convenient metric to quantify the sensitivity of these observables to constrain the parameters of our model is given by the Fisher information~\cite{Brehmer:2016nyr, Brehmer:2017lrt}. Its component describing the sensitivity to the CP-phase $\alpha$ is defined as
\be
I= \mathbb{E} \left[
\frac{\partial \log p(x|\kappa_t, \alpha)}{d \alpha}
\frac{\partial \log p(x|\kappa_t, \alpha)}{d \alpha}
\right] \ . 
\label{eq:info}
\ee
Here, $p(x|\kappa_t, \alpha)$ is the likelihood function, which describes the probability to observe a set of events with corresponding observables $x$ as a function of the model parameter $\kappa_t$ and $\alpha$. $\mathbb{E} [\cdot]$ denotes the expectation value evaluated at the SM point, $(\kappa_t,\alpha)_{\rm SM}=(1,0)$. In the following we use the \texttt{MadMiner} package to calculate the Fisher information~\cite{Brehmer:2019xox}.

In the left panel of \cref{fig:information}, we show the Fisher information associated with the CP-sensitive spin correlation observables for the di-leptonic (red), semi-leptonic (gray), and hadronic (blue) channels. The bars on the left show the \textit{full} information, \emph{i.e.}, the information that could be accessed via a comprehensive multivariate analysis. This was estimated using the machine learning method based on the \texttt{SALLY} algorithm~\cite{Brehmer:2018eca, Brehmer:2018kdj, Brehmer:2018hga} trained with all possible spin correlation observables. The remaining bars show the information in individual observables $\Delta \phi_{ik}^{t\bar t}$, which were estimated using a histogram based approach. 

Focusing first on the di-leptonic channel, the most sensitive among these observables is the spin correlation between the leptons, $\Delta\phi^{t\bar t}_{\ell\ell}$, since the spin analyzing power for the charged leptons are maximal. The next most sensitive observables are those where a charged lepton has been replaced with a $b$-jet or a $W$ boson. We observe that the corresponding Fisher information in $\Delta\phi^{t\bar t}_{\ell b}$ and $\Delta\phi^{t\bar t}_{\ell W}$ are suppressed compared to $\Delta\phi^{t\bar t}_{\ell\ell}$ by the square of the spin analyzing power $\beta_{b/W}^{2} \sim 0.4^{2}$, as expected. The information in the spin correlations observables between a pair of $b$-jet(s) and/or $W$ boson(s) is further suppressed by an additional factor of $\beta_{b/W}^{2}$.

Let us now also consider the other top decay channels. As the Fisher information is proportional to the rate~\cite{Brehmer:2016nyr}, we expect it to increase relative to the fully leptonic channel by a factor ${2\times\text{BR}_{W \!\to\text{had}}/\text{BR}_{W \!\to\text{lep}} \sim 6}$ for the semi-leptonic channel and $(\text{BR}_{W \!\to\text{had}} /\text{BR}_{W \!\to\text{lep}})^2\sim 9$ for the hadronic channel. Looking at the last three observables involving $b$-jets and $W$-bosons, this is indeed the case. For the other observables, we notice an additional loss of about a factor 2 in spin analyzing power, and hence a factor 4 in the Fisher information, which is caused by probing $j_\text{soft}$ instead of the $d$-quark. 

Overall, we see that the different observables have distinct overall importance in the three channels. For di-leptonic top decays, most of the information is contained in the spin correlation between the leptons, while the information in other observables is significantly suppressed. In contrast, for the hadronic decay channel, all shown observables have almost similar information. In this case, the resulting full information, that can be obtained by combining the different spin correlation observables, significantly exceeds the information of any individual observable. Overall, all three channels carry a similar amount of information on the CP-phase $\alpha$, which suggest performing a combined analysis. \medskip

Due to the limited $t\bar{t}h$ event rate at the LHC, we expect the non-linear new physics effects to dominate over the linear ones. We therefore expect most of the sensitivity on the CP-structure of the top Yukawa coupling to arise from these non-linear terms, despite the fact that the corresponding observables are not genuine CP-sensitive. To quantify the sensitivity of these CP-even observables to the squared terms, we use modified version of the Fisher information that was introduced in Ref.~\cite{Brehmer:2019gmn}. In this approach, we simply consider the square of the coupling as our new model parameter and define
\be
I' = \mathbb{E} \left[
\frac{\partial \log p(x|\kappa_t^2, \alpha^2)}{d \alpha^2}
\frac{\partial \log p(x|\kappa_t^2, \alpha^2)}{d \alpha^2}
\right] \ . 
\ee
The result is shown in the right panel of \cref{fig:information}. Here, we show the information associated with a two-dimensional distribution of two observables, relative to the full information associated with a multivariate analysis using all observables. As none of the presented observables relies on the top quark final state kinematics, the results are identical for all three top quark decay channels. 

The distribution of the invariant mass of the photon pair, $m_{\gamma\gamma}$, is only sensitive to the theory parameters through its normalization. Correlating it with itself, we obtain the information associated with the signal strength measurements, which accounts for $31\%$ of the information on the CP-phase. In the absence of background, the correlation of $m_{\gamma\gamma}$ and any other observable is equivalent to the information in a single differential distribution of that observable. This is shown in the bottom row. As expected, it is also identical to the information for the correlation of an observable with itself, which are shown in the diagonal. We can identify $\Delta \eta_{t\bar t}$ and $\theta^*$ as the two most sensitive observables, which individually carry about $60\%$ of the full information. 

Combing two different observables further increases the information. The two most promising combinations are $\Delta \eta_{t\bar t}$ vs. $p_{Th}$ as well as $\theta^*$ vs. $m_{t\bar t h }$, which carry about $73\%$ of the full information. Successively adding more observables further increases the information. This shows that a multivariate analysis is vital to maximize the sensitivity on the CP-phase $\alpha$.

\section{Kinematic Reconstruction}
\label{sec:kin_reconstruct}

Most of the new physics probes discussed so far, $viz.$ $m_{tt}$, $\theta^{*}$, $b_{4}$, and $\Delta \phi_{ik}^{t\bar{t}}$, require a full reconstruction of the top and anti-top momenta. This is a challenging task at the LHC due to combinatorial ambiguities and the  presence of up to two neutrinos in the $t\bar{t}(h\to \gamma\gamma)$ final state. In this section, we discuss the strategies adopted for the kinematic reconstruction of the semi-leptonic and hadronic channels, and the more complex di-leptonic mode. 

\medskip
\noindent\textbf{Semi-leptonic channel}: In the semi-leptonic channel, the full reconstruction of the $t\bar{t}$ system requires the determination of the longitudinal momentum of the missing particle $\nu$. We compute it by constraining the invariant mass of the lepton and the neutrino to the $W$-boson mass. Typically, either two solutions or zero solutions are obtained. Around $35\%$ of events give zero solutions, and discarding all such events would lead to a significant reduction in event statistics. Therefore, in such events, we vary the transverse momentum of the missing system~(at most by $\pm 10\%$) while keeping the azimuth angle of $\nu$ unchanged until physical solutions of $p_{z,\nu}$ are obtained. Events which give zero solutions even after the aforesaid variation are ignored. 
We perform the reconstruction for the top quarks iterating over all possible partitions of light jets~($j$) and $b$-jet forming the hadronic top~($jjb$) and leptons and $b$-jet for the leptonic top~($\ell \nu b$). The two possible neutrino solutions are separately accounted for, forming different partitions. We select the combination that minimizes 
\begin{align}
(m_{jjb}-m_t)^2+(m_{\ell\nu b} - m_t)^2,
\end{align}
where $m_{t}$ is the on-shell mass of the top quark. 

\medskip
\noindent\textbf{Hadronic channel:} We follow a similar mass minimization strategy in the hadronic channel. We reconstruct the two top quarks, $t_1$ and $t_2$, by iterating over all possible combinations of light jets and $b$-jets.  
The combination which minimizes  
\begin{align}
(m_{t_1}-m_t)^2+(m_{t_2} - m_t)^2,
\end{align}
is chosen. 

\medskip
\noindent\textbf{Di-leptonic channel}: In the more complex di-leptonic $t\bar{t}h$ channel, the invisible system is constituted by two neutrinos. Therefore, in addition to determining the unknown longitudinal momentum of the missing particles, it is also indispensable to partition the four-momentum of the missing system into the two neutrinos in order to fully reconstruct the top and the anti-top. An additional combinatorial ambiguity arises from the tandem $b$-jet  and $\ell$ pairing. The study  in Ref.~\cite{Goncalves:2018agy} reconstructed the $t\bar{t}(h\to b\bar{b})$ system in di-leptonic mode using $M_{2}$ assisted reconstruction algorithm and a boosted $h\to b\bar{b}$, with jet substructure techniques, to suppress the additional combinatorics between the Higgs boson and top quark decays.
In contrast, the present analysis reconstructs the $t\bar{t}(h\to \gamma\gamma)$ system following the Recursive Jigsaw Reconstruction~(RJR) algorithm presented in Ref.~\cite{Jackson:2017gcy}.
The RJR approach utilizes a series of jigsaw rules optimized to estimate the unknown kinematic degrees of freedom in an event topology and resolve the combinatorial ambiguities between/within the final state visible and invisible objects. It results is a complete kinematic basis which can be used to define the four-momenta of all the final state and intermediate objects in an event decay tree. 

The first step involves the resolution of combinatorial ambiguity between the $b$-jets and the leptons by using the ``Combinatorial Minimization" Jigsaw Rule~(JR)~\cite{Jackson:2017gcy}, identifying the ($b$-jet, $\ell$) pairs by minimizing
\begin{equation}
        (m_{ b_j \ell^{+}}^{2} + m_{b_k \ell^{-}}^{2});~ j,k=1,2;~ j \neq k. 
    \end{equation}

After establishing the two visible hemispheres corresponding to the top and the anti-top, we apply the ``Invisible Mass" JR to estimate the invariant mass of the invisible system~($m_{I}$)~\cite{Jackson:2017gcy} defined as
\begin{equation}
    m_{I}^{2} = m_{V}^{2} - 4m_{V_{a}}^{2}m_{V_{b}}^{2},
\end{equation}
where $m_{V}$ is the invariant mass of all the two $b$-tagged jets and the two leptons in the final state. $m_{V_{a}}$ and $m_{V_{b}}$ correspond to the invariant mass of the two visible hemispheres associated with the top and the anti-top that were reconstructed in the previous step. $m_{I}$ is chosen such that it is the smallest Lorentz invariant mass that ensures a non-tachyonic four-momenta for the individual neutrinos upon partitioning the invisible system. Next, we determine the longitudinal momentum of the invisible system, $\slashed{p}_{z}$, using the following relation given by the ``Invisible Rapidity" JR~\cite{Jackson:2017gcy}:
\begin{equation}
    \slashed{p}_{z} = p_{z}^{V} \frac{\sqrt{{|\slashed{p}_{T}|}^{2} + m_{I}^{2}}}{\sqrt{{|p_{T}^{V}|}^{2} + m_{V}^{2}}}.
\end{equation}
Here, $p_{z}^{V}$ and $p_{T}^{V}$ represent the longitudinal and transverse momenta, respectively, of the visible system constituted by the two $b$-jets and the two leptons, while $\slashed{p}_{T}$ is the missing transverse momentum. 

At this point, we have all the ingredients required to reconstruct the $t\bar{t}$ system. However, in order to reconstruct the top and the anti-top individually, the invisible four-momentum has to be correctly partitioned into the two neutrinos. This is achieved by using the ``Contraboost Invariant" JR specified in Ref.~\cite{Jackson:2017gcy} which estimates the four-momenta of the neutrinos produced from top and anti-top decay in the $t\bar{t}$ CM frame under the assumption that both $t$ and $\bar{t}$ have the same invariant mass. The resolved four-momenta of the neutrinos along with the correctly paired $b$-jets and leptons allows defining the $t$ and the $\bar{t}$ systems independently. The reconstruction efficiency of this method is about $80\%$, which is comparable with $M_{2}$ assisted reconstruction algorithm~\cite{Goncalves:2018agy}.

\medskip
With the fully-resolved $t\bar{t}h$ system, we can reconstruct a multitude of CP-even and CP-odd spin correlation observables defined in the $t\bar{t}$ CM frame and the lab frame. Several observables that do not depend on the spin-polarization of $t\bar{t}$ pair are also considered. Our goal here is to maximally explore the $t\bar{t}h$ multi-particle final state, augmenting the CP-sensitivity of the $pp \to t\bar{t}h$ channel at the HL-LHC. 

\section{Analysis}
\label{sec:analysis}

\subsection{Simulation and Event Selection}

\begin{figure*}[!tb]
    \centering
    \includegraphics[width=0.32\textwidth]{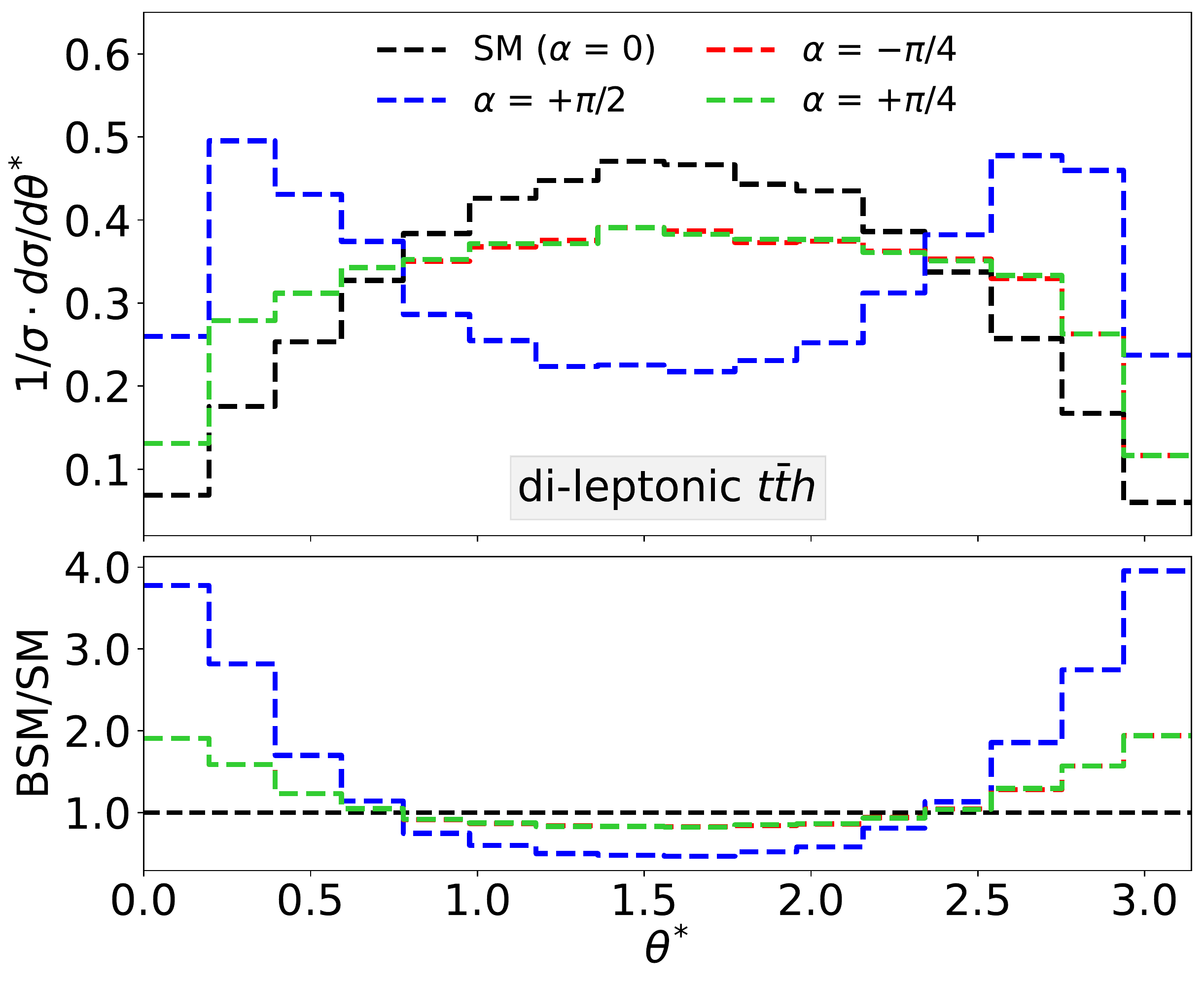}
    \includegraphics[width=0.32\textwidth]{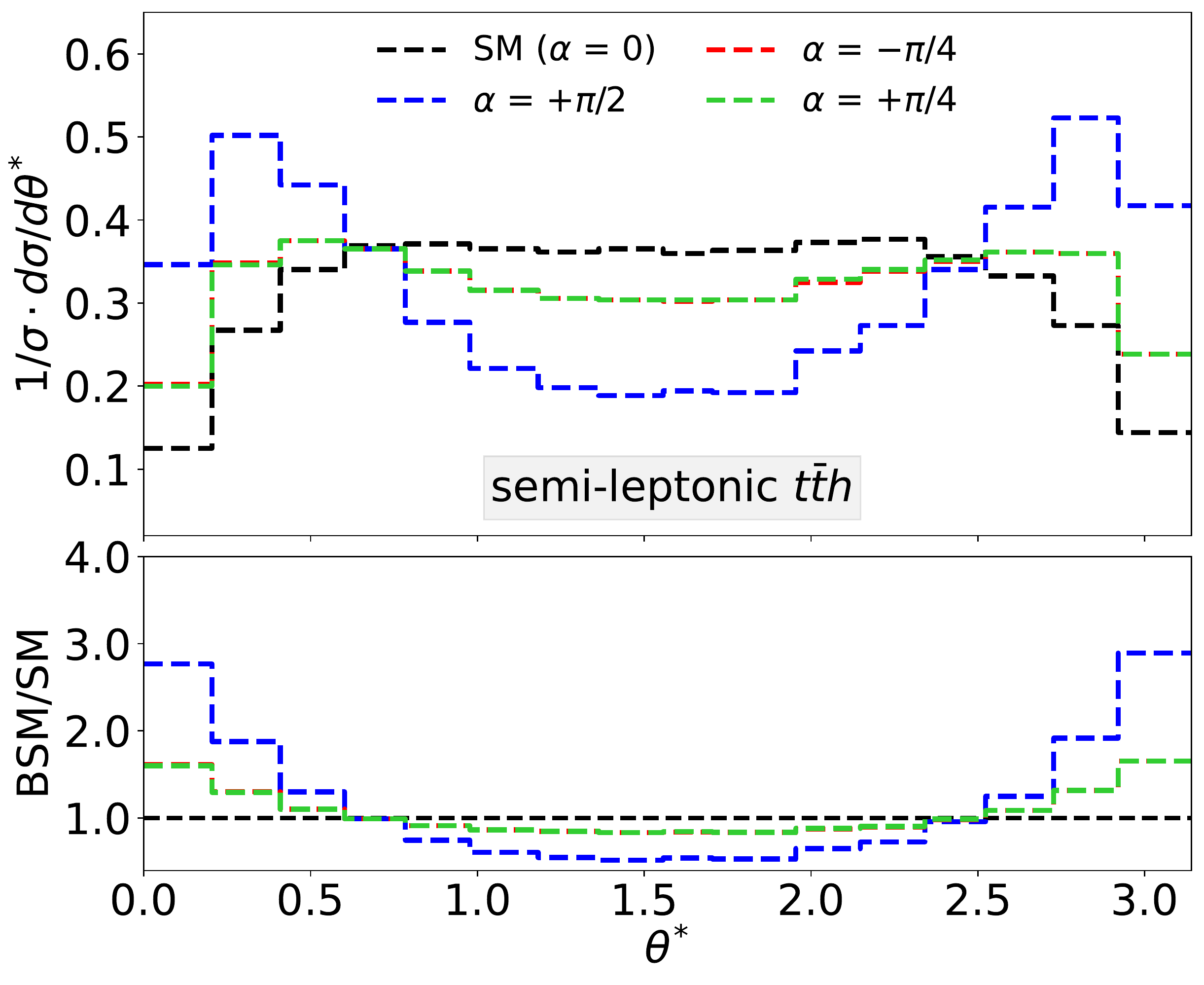}
    \includegraphics[width=0.32\textwidth]{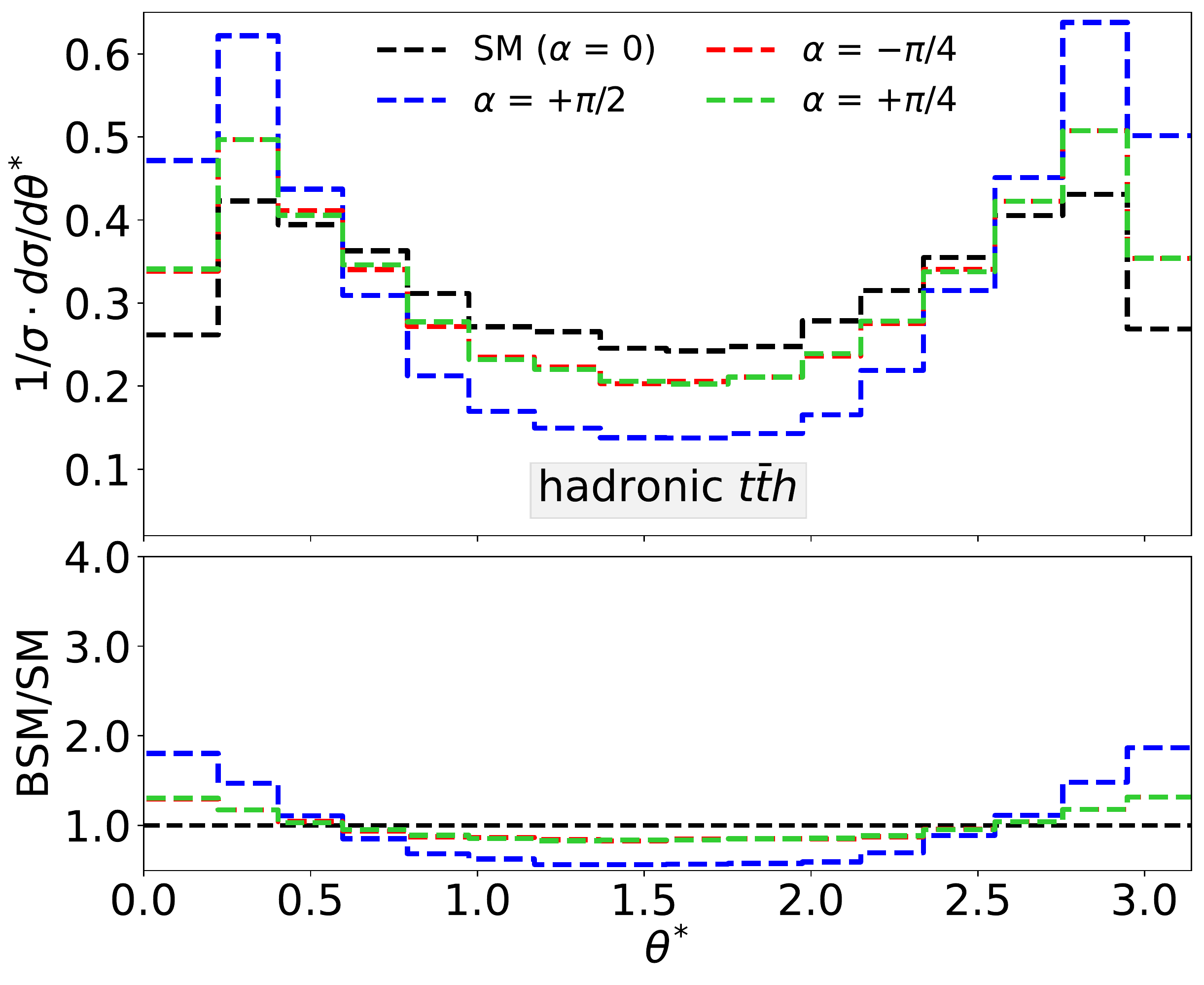}
    \caption{Reconstructed detector level distributions for the Collins-Soper angle $\theta^{*}$ for the di-leptonic channel (left),  semi-leptonic channel (center), and  hadronic channel (right). We present for the SM $(\alpha=0)$ and several beyond the SM Higgs-top CP-hypotheses $(\alpha=\pi/2,\pm \pi/4)$.}
    \label{fig:theta-star_delphes}
\end{figure*}

In this section, we explore the \emph{direct} Higgs-top CP measurement combining machine learning techniques and efficient kinematic reconstruction methods. We consider $t\bar{t}h$  signal with $h \to \gamma\gamma$ in the di-leptonic, semi-leptonic, and hadronic top decay modes at the HL-LHC. The dominant background to this process is given by continuum $t\bar{t}\gamma\gamma$ production. We simulate both the signal and background event samples with \texttt{MadGraph5$\_$aMC@NLO}~\cite{ Alwall:2014hca} within the \texttt{MadMiner} framework~\cite{Brehmer:2019xox} at leading order~(LO) with a center-of-mass energy of $\sqrt{s} = 14~\tev$. Higher order effects to the signal rate are included via a flat next-to-leading order k-factor~\cite{deFlorian:2016spz, HXSWG}.
We use \texttt{NNPDF2.3QED} parton distribution function~\cite{Ball:2013hta}. No generation-level cuts have been applied for the signal events, while the backgrounds have been generated in the mass window $105~\gev \!<\! m_{\gamma\gamma} \!<\! 145~\gev$. Parton shower and hadronization effects have been included with \texttt{Pythia~8}~\cite{Sjostrand:2007gs} and fast detector simulation with the \texttt{Delphes3} package~\cite{deFavereau:2013fsa}, using the default HL-LHC detector card~\cite{HLLHC_card,Cepeda:2019klc}.

To obtain the cross section and likelihood function as a function of the theory parameters, we use the morphing technique that is already implemented in \texttt{MadMiner}. Here, we take into account the dependence of new physics theory parameters at both $t\bar{t}h$ production and $h \to \gamma\gamma$ decay, and therefore choose a quartic ansatz in the morphing setup, which is used to interpolate the event weights as a function of $\kappa_H=\kappa_{t} \cos{\alpha}$ and $\kappa_A= \kappa_t \sin\alpha$.

We start our analysis by selecting events consisting of two photons and at least two $b$-tagged jets. In addition, we require the final state to contain exactly two opposite-sign leptons for the di-leptonic channel, exactly one lepton and at least two light jets for the semi-leptonic channel, and at least four light jets for the hadronic channel. We demand the individual particles to pass the following identification cuts: 
\be
    p_{T\ell} &\!>\! 15~\gev,   \      |\eta_\ell| \!<\! 4,
    \quad 
    p_{T\gamma} \!> \!15~\gev,  \      |\eta_\gamma| \!<\! 4,
    \\
    p_{Tb} &\!>\! 25~\gev,      \      |\eta_b| \!<\! 4,    
    \quad 
    p_{Tj} \!>\! 25~\gev,       \      |\eta_j| \!<\! 5.
\ee
In addition, we require the di-photon invariant mass to satisfy $|m_{\gamma\gamma} \!-\! 125| \!<\! 10~\gev$. 

We fully reconstruct the $t\bar{t}h$ system following the strategy described in \cref{sec:kin_reconstruct}. In particular, this allows to obtain both the lab frame and the $t\bar{t}$ CM frame observables. As an example for an observable that requires the top reconstruction, we present the distribution of the Collins-Soper angle $\theta^{*}$ in \cref{fig:theta-star_delphes}.  When comparing these detector level distributions to the result at parton level, presented in \cref{fig:semilep_parton_dist}, we observe the robustness of our analysis in respect to the reconstruction strategy and detector effects. The distributions are found to retain the CP sensitivity at the detector level, albeit a reduction of about $20\%$ for the di-leptonic channel, $40\%$ for the semi-leptonic channel and $50\%$ for the hadronic channel, compared to parton level. 
 
\subsection{Analysis Methodology}

As we have seen in \cref{sec:theory}, there is no single observable that carries all the information on the CP-structure of the top quark Yukawa. Instead, there is a variety of sensitive observables. Hence, a multi-variate analysis is needed to extract the maximal information on the theory parameters from the data. In the following, we will summarize the adopted observables and the analysis methodology. 

In this analysis, we consider the following list of 80 observables to describe the kinematics of signal and background events.
\be
   \textbf{Observables: } & 
   \Delta\phi_{ik}^{t\bar{t}}, \ 
   \Delta\phi_{hi(k)}^{t\bar{t}}, \ 
   \Delta \phi_{ik}^\text{lab},\\
    & 
    \ 
   \Delta \phi_{h t/h\bar{t}},  \ 
   \Delta\phi_{t\bar{t}},  \theta^{*}, \ 
   b_{4},\  \\ 
         & 
   m_{\gamma\gamma}, \ 
   m_{t\bar{t}}, \ 
   m_{t\bar{t}h},\Delta R^{\rm 2^{nd}min}_{\gamma j},\\
   & 
   \Delta R_{j_\text{soft}j_\text{hard}}, \  
   \Delta R_{\ell\nu},\Delta R_{Wb}, \ 
   \Delta R^\text{min}_{\gamma j}, \\ & 
   \Delta \eta_{t\bar{t}}, \ 
   m_{ht/h\bar{t}}, \ 
   H_{T},  
   \slashed{E}_{T}/\sqrt{H_{T}}, \\ 
 & 
   \{p_{T},~\eta\}_{X} \text{ for } X =i, k, t, \bar{t},h .
\ee
We include the complete set of observables used by the ATLAS collaboration in a recent Higgs-top CP study~\cite{ATLAS:2020ior} and complement this set with additional CP-even observables that show strong sensitivity to the CP-phase  $(\theta^*, b_4,m_{t\bar{t}},  m_{t\bar{t}h})$ together with the transverse momentum and pseudorapidity of all final state and reconstructed objects. We also incorporate a comprehensive list of spin correlations, as introduced in \cref{eq:CP_odd_obseravable}, which are constructed between all possible final state pairs. We include both observables constructed in the $t\bar{t}$ rest frame, $\Delta\phi_{ik}^{t\bar{t}}$, and in the lab frame, $\Delta\phi_{ik}^{\text{lab}}$. Finally, we  account for the correlation observables $\Delta\phi_{hi(k)}^{t\bar{t}}$  that arise from the tensor products involving the Higgs boson momentum, $\epsilon (p_t,p_{\bar t}, p_h, p_{i(k)})$. The following pairs $\{i,k\}$ are considered for the different channels

\be 
    \text{di-leptonic: }
    i &= \ell^{+},\nu_{t} ,b_{t}, W_{t} \\
    k &= \ell^{-},\nu_{\bar{t}} ,b_{\bar{t}}, W_{\bar{t}} \\
    \text{semi-leptonic: }
    i &= {\ell,\nu,b_{\ell},W_{\ell}} \\
    k &= {j_\text{soft},j_\text{hard}, b_\text{had}, W_\text{had}}\\
    \text{hadronic: }
    i &= j_\text{soft}^{t_1}, j_\text{hard}^{t_1}, b_{t_1}, W_{t_1} \\
    k &= j_\text{soft}^{t_2}, j_\text{hard}^{t_2}, b_{t_2}, W_{t_2}
\ee
In the semi-leptonic case, $b_{\ell}/W_{\ell}$ and $b_\text{had}/W_\text{had}$ represent the $b$-jets/$W$-bosons produced from the leptonically and hadronically decaying top quarks, respectively. In events with more than two $b$-tagged jets, the hardest two are considered while reconstructing the top and the anti-top quarks. $j_\text{hard}$ corresponds to the hardest light jet, from the hadronic top quark, in the top rest frame. \medskip 

To interpret the results of our analysis and obtain projected sensitivities, we follow a likelihood-based approach. According to the Neyman-Pearson lemma, the most powerful test statistic to discriminate between two hypotheses, in our case a new physics model parameterized by $\theta = (\kappa_t,\alpha)$ and the SM with $\theta_\text{SM} = (1,0)$, is the likelihood ratio $r(x|\theta; \theta_\text{SM})$. Here, $x$ denotes the set of reconstructed observables introduced above. 

\begin{figure*}
    \centering
    \includegraphics[scale=0.34]{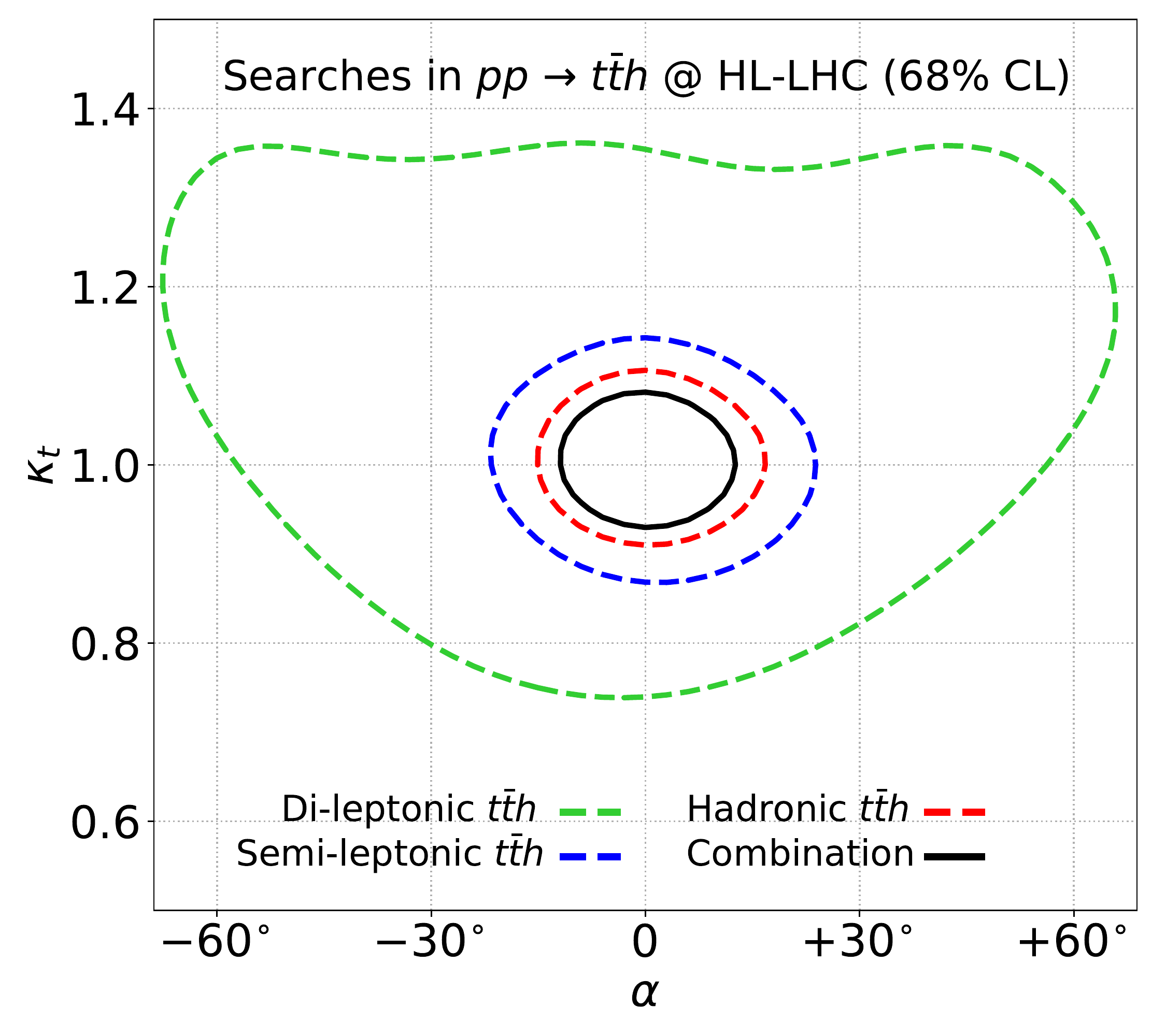}\hspace{1cm}
    \includegraphics[scale=0.34]{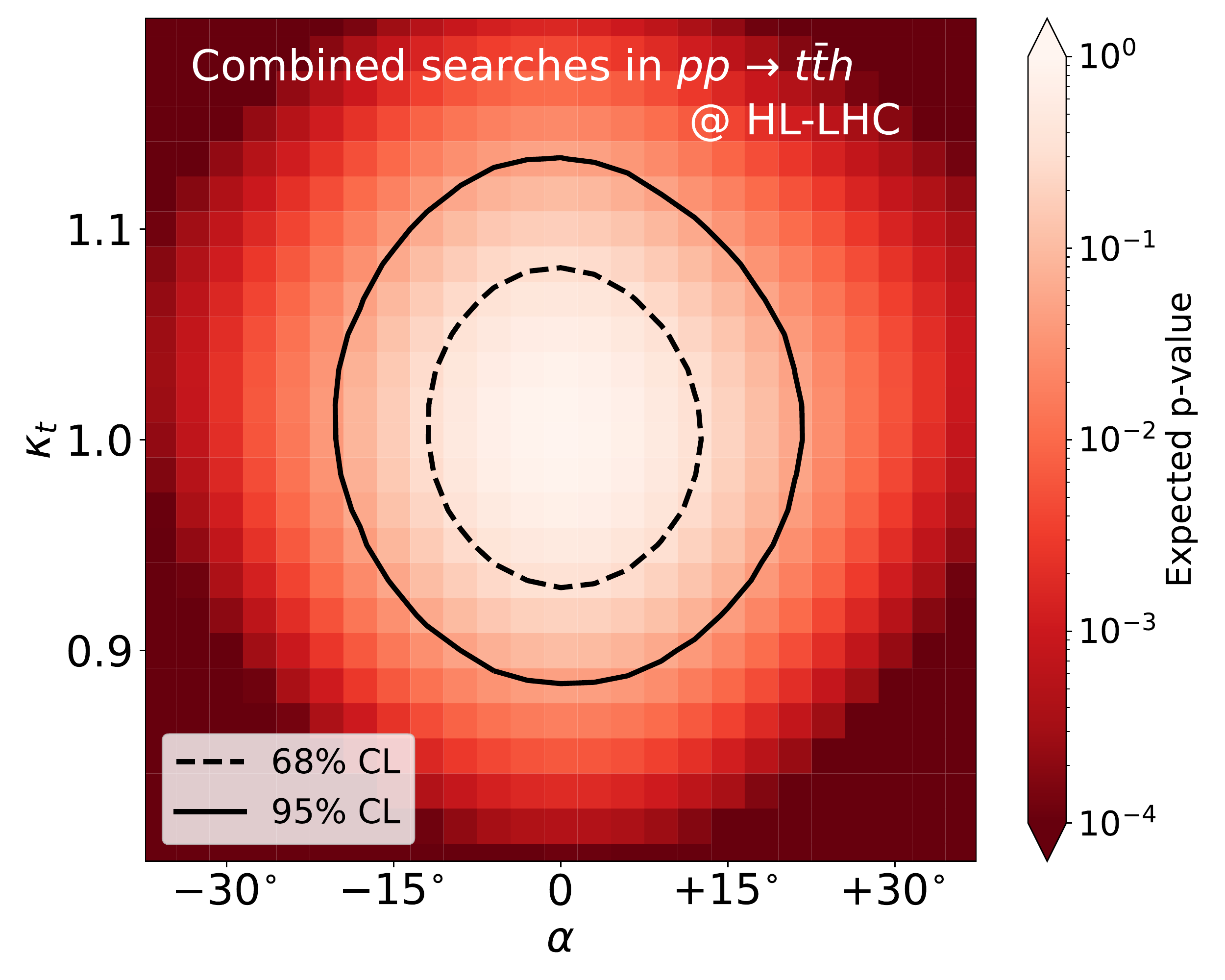}
    \caption{Projected sensitivity in the $(\alpha,\kappa_t)$ plane. In the left panel we show the projected $68\%$ CL contours from direct Higgs-top searches in the semi-leptonic~(blue), di-leptonic~(green), hadronic~(red) $t\bar{t}h$ channels, and their combination~(black), considering all input observables. The right panel shows the projected $68\%$~CL (dashed) and $95\%$~CL (solid) contours from the combination of the three channels considering all input observables. The color palette illustrates the expected p-value of the estimated log-likelihood ratio. The projections are derived for 14~TeV LHC assuming an integrated luminosity of $3~\iab$.} 
    \label{fig:combined_excl_limits}
\end{figure*}

Whereas the likelihood ratio involving detector level observables is intractable, meaning that it cannot be computed directly, it can be estimated using simulations. To address this issue, we use the machine learning based technique introduced in Refs.~\cite{Brehmer:2018eca, Brehmer:2018kdj, Brehmer:2018hga, Stoye:2018ovl, Brehmer:2019bvj, Brehmer:2020ako, Brehmer:2020zwh}, which has been implemented in the \texttt{MadMiner} tool~\cite{Brehmer:2019xox}. This approach uses both reconstructed observables and matrix-element information, which are then used to train neural network models that estimate the likelihood ratio. It therefore accounts for the effects of parton shower, hadronization, and detector effects, while the matrix-element information helps to significantly improve the performance of the neural network training.  Using the estimated likelihood ratio function $r(x|\theta; \theta_\text{SM})$, which describes both the linear and non-linear new physics effects, we then perform a likelihood ratio test to obtain our projected sensitivities. 

We simulate $10^6$ signal and $10^6$ background events before event selection. Using \texttt{MadMiner}, we train neural networks to estimate the likelihood ratio using the \texttt{ALICES} algorithm with its hyperparameter set to unity~\cite{Stoye:2018ovl}. We use fully connected neural networks with three hidden layers, each containing 100 nodes and $\tanh$ activation function. The neural network training is performed over 100 epochs using the Adam optimizer. To avoid overtraining, we evaluate the loss function on an independent validation set and employ an early stopping procedure. We use a batch size of 128, and an exponentially decaying learning rate~(from $10^{-4}$ to $10^{-5}$). The limit setting is performed with \texttt{MadMiner}'s \texttt{Likelihood} class. 

\subsection{Results}

Let us now turn to the results of our study. In \cref{fig:combined_excl_limits} we show the projected sensitivity on the top Yukawa coupling in terms of $\kappa_t$ and $\alpha$ using the $t\bar{t}(h\to \gamma\gamma)$ measurement. In the left panel, we present the $68\%$ CL contours for the individual top decay channels as colored dashed lines. A combination of  all channels is shown in the black solid line. The studied channels can be organized in ascending order of sensitivity as: di-leptonic, semi-leptonic, and hadronic modes. Since the leading observables display efficient reconstruction for all channels, as illustrated in Fig.~\ref{fig:theta-star_delphes}, the order of sensitivity among the final state modes follow their correspondent event rate.

In the right panel, we show the $68\%$ and $95\%$ CL contours as dashed and solid lines, respectively. The p-values in the $({\kappa_t,\alpha})$ parameter space are presented through the color palette. We observe that $|\kappa_t|$ can be constrained within $\mathcal{O}(8\%)$ of the SM value at $68\%$ CL through a combination of direct searches in the $t\bar{t}(h\to\gamma\gamma)$ channel at the HL-LHC. Assuming $\kappa_t = 1$, the combined search would be able to probe the Higgs-top CP phase up to $|\alpha| \lesssim 13^{\circ}$ at $68\%$ CL. 

We also perform a separate analysis in which we train a neural network exclusively with the CP-even observables shown in the right panel of \cref{fig:information}. We observe that the projected sensitivity of such an analysis, using this smaller set of CP-even observables that are most-sensitive to the non-linear new physics effects, is almost comparable to the projected sensitivity of the combination study performed using the full set of observables. Overall, almost all the sensitivity to the Higgs-top CP-structure is provided by the non-linear terms in $\alpha$. The limited $t\bar{t}(h\to\gamma\gamma)$ event statistics renders sub-leading sensitivity to the observables which probe the linear terms. 

\subsection{Systematic Effects}

In this section, we explore the implications from systematic uncertainties on the projected sensitivity of $\kappa_{t}$ and $\alpha$. In particular, we will consider two sources of uncertainty associated with the normalization of both signal and background. 

In the statistical analysis, these uncertainties are parameterized through nuisance parameters $\nu_{s}$ and $\nu_{b}$ for the signal and background normalization, respectively. These nuisance parameters encode theoretical and experimental uncertainties on the normalization of  distributions, neglecting possible shape uncertainties. As before, we train a neural network using the \texttt{ALICES} method in \texttt{MadMiner} to estimate the likelihood ratio $r\left(x | \theta, \nu; \theta_{SM}, \nu_{SM} \right)$. This is now a function of both the model parameters $\theta = (\kappa_t, \alpha)$ and the nuisance parameters $\nu = (\nu_{s},\nu_{b})$ which have a nominal value $\nu_{SM} = (0,0)$. Before setting limits, a constraint term describing our prior knowledge on the nuisance parameter is added. Adopting a conservative approach, we assume a prior constraint of $20\%$ and $50\%$ in the $t\bar{t}(h \to \gamma\gamma)$ signal and the $t\bar{t}\gamma\gamma$ background, respectively. Finally, we profile over the nuisance parameters following the procedure described in Ref.~\cite{Brehmer:2018eca}. 

Before turning to the sensitivity contours, let us remind ourselves that the presented results are based on a multi-variate analysis. In particular, this includes the invariant mass of the di-photon pair. The considered range, $115~\gev < m_{\gamma\gamma} < 135~\gev$, was chosen sufficiently wide to contain both a signal dominated region at the Higgs resonance and a background dominated region around it. \texttt{MadMiner} uses this background dominated region to constrain the nuisance parameter associated with the background normalization $\nu_b$, and therefore effectively performs a data-driven side-band analysis. As we will see in a moment, the effective uncertainty of the background normalization is therefore significantly smaller than the 50\% which we assumed as a prior. 

\begin{figure}[t]
    \centering
    \includegraphics[scale=0.37,trim={5mm 0 0 0},clip]{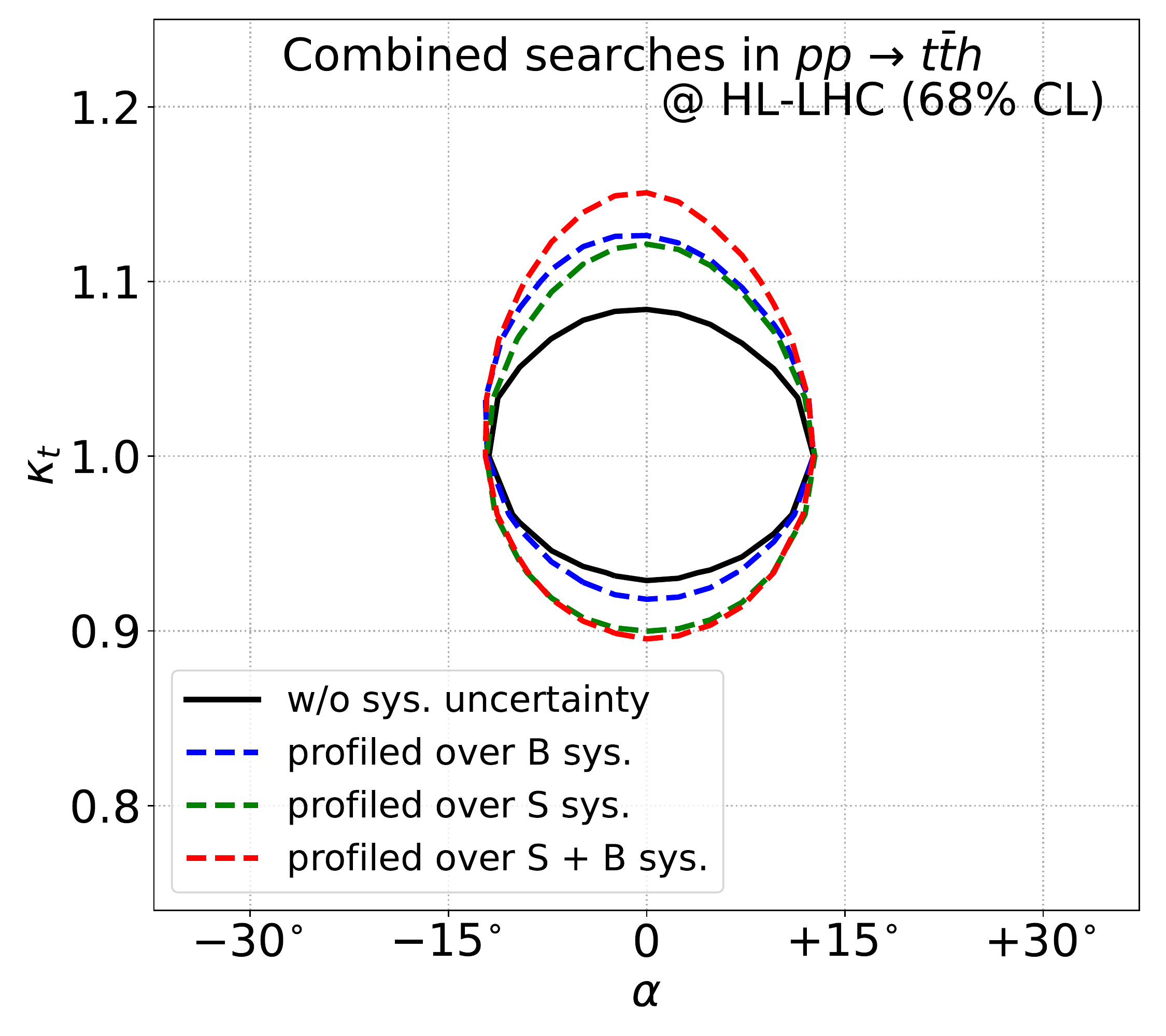}
    \caption{Projected $68\%$ CL contours on $(\alpha,\kappa_{t})$ from the combination of the three channels considering all input observables, profiled over background~(B) uncertainties~(blue), signal~(S) uncertainties~(green), and both signal and background systematic uncertainties~(red). The projections are derived for 14 TeV LHC assuming $\mathcal{L} = 3~\iab$.}
    \label{fig:combined_systematics_excl}
\end{figure}

In the following, we analyze three scenarios to study the impact of systematic uncertainties on the projected sensitivity in the $\left(\alpha, \kappa_{t}\right)$ plane. In the first scenario, we study the impact associated with only the uncertainty on the background normalization. To do so, we fix the $\nu_{s}=0$ in the estimated likelihood ratio and profile over $\nu_{b}$. Similarly, in a second scenario, we fix $\nu_{b}=0$ and we profile over $\nu_{s}$ to study the impact of the signal uncertainty. Finally, in a third scenario, we obtain the limits after profiling the likelihood ratio over both $\nu_{s}$ and $\nu_{b}$. In \cref{fig:combined_systematics_excl} we present the projected sensitivity on $\alpha$ and $\kappa_{t}$ for all scenarios. The blue, green, and red contours correspond to the first, second, and third scenarios, respectively. The black contour represents the sensitivity assuming no systematic uncertainty and corresponds to the black-solid contours in \cref{fig:combined_excl_limits}.

At first, we observe that the sensitivity in $\alpha$ remains unaffected from systematic uncertainties~\cite{Goncalves:2021dcu}. This stems from the reason that at $\kappa_{t} = 1$, the sensitivity in $\alpha$ is dominantly controlled by the shape information from kinematic distributions and is largely independent of the event rate due to the combination of two competing effects. On the one hand, the signal cross section $\sigma_{t\bar{t}\left(h \to \gamma\gamma\right)}$ decreases with $\alpha$: for example at $\kappa_t = 1$ the cross section $\sigma_{t\bar{t}\left(h \to \gamma\gamma\right)}$ falls by $\mathcal{O}(25\%)$ from $\alpha = 0$ until $\alpha \sim \pi/3$ and then remains roughly unchanged until $\alpha = \pi/2$. On the other hand, the signal efficiency also improves with $\alpha$. These two effects roughly offset any overall dependence on the event rate, thereby leading to unchanged projection contours in the direction of $\alpha$ even after profiling over the nuisance parameters. 

The situation is qualitatively different in the $\kappa_t$ direction. When $\alpha = 0$, the measurement is purely based on a rate information, implying that the Higgs coupling strength $\kappa_t$ and the signal normalization, as parameterized by $\nu_s$, are essentially degenerate. Therefore, our prior uncertainty of the signal normalization will directly propagate into a systematic uncertainty on $\kappa_t$. The new physics effects in the Higgs-top coupling manifest as $\sim \kappa_{t}^{2}$ at the $t\bar{t}h$ production level and as $ \sim \left(1.28 - 0.28\kappa_{t}\right)^{2}$ in $h \to \gamma\gamma$ decay~\cite{Brod:2013cka}. After combining these two factors, an uncertainty of $20\%$ in the $pp \to t\bar{t}(h \to \gamma\gamma)$ cross section translates to roughly $12\%$ uncertainty in $\kappa_{t}$. We observe this effect in \cref{fig:combined_systematics_excl}: for $\alpha = 0$ the projected sensitivity falls from $|\kappa_{t}|\lesssim 8\%$ in the absence of systematic uncertainties to $|\kappa_{t}| \lesssim 13\%$ on profiling over $\nu_{s}$. We observe that despite a prior $50\%$ uncertainty in the background normalization compared to $20\%$ in the signal, its impact on the projection contours in the $\left(\kappa_t, \alpha \right)$ plane is milder. As discussed above, this is a consequence of the side-band measurement and illustrates the robustness of our multi-variate analysis.     



\section{Summary}
\label{sec:summary}

In this study, we derived the prospects to \emph{direct} measure the Higgs-top CP-structure in $t\bar{t}(h\to \gamma\gamma)$ channel at the HL-LHC. We show that a combination of machine learning techniques and efficient kinematic reconstruction methods can boost  new physics sensitivity, effectively exploring the complex $t\bar{t}h$ multi-particle phase space. 

Among the several probes included in our machine learning analysis, this study encompass a comprehensive set of spin correlation observables. Beyond the SM CP-phases steer the spin-polarization of the top pair, and the spin correlations are carried forward by their decay products. We harness the potential of the spin correlation observables via the full reconstruction of the top and anti-top, evaluating these particular observables in the $t\bar{t}$ CM frame, where the correlations are maximal. In the hadronic and semi-leptonic $t\bar{t}h$ channels, we used mass minimization to fully reconstruct the $t\bar{t}h$ system. In the more complex di-leptonic channel, we employed the Recursive Jigsaw Reconstruction technique to resolve the combinatorial ambiguities and determine the unknown degrees of freedom. In all channels, the effects of parton showering, hadronization, and detector resolution were included.

Exploring the intricate $t\bar{t}h$ multi-particle phase space  with CP-odd and even observables defined in the laboratory frame and the $t\bar{t}$ CM frame, we obtain strong projections for the Higgs-top CP-phase. Through a combined semi-leptonic, hadronic, and di-leptonic $t\bar{t}(h\to\gamma\gamma)$ search, the HL-LHC can directly probe the Higgs-top coupling modifier and CP-phase respectively up to $|\kappa_t|\lesssim 8\%$ and $|\alpha|\lesssim 13^{\circ}$ at $68\%$~CL.


Possible improvements can be expected by including other relevant channels, such as $t\bar{t}(h\to b\bar{b})$~\cite{Buckley:2015vsa,AmorDosSantos:2017ayi,Goncalves:2018agy,Goncalves:2021dcu}. While this channel displays the bulk of the Higgs decay, $\mathcal{BR}(h\to b\bar{b})\sim 58\%$, it results in sub-leading limits in comparison to $t\bar{t}(h\to \gamma\gamma)$ as it  endures a substantial QCD background that is associated with sizable uncertainties~\cite{ATLAS:2017fak,CMS:2018hnq}. Fast-moving precision calculations~\cite{Jezo:2018yaf,Denner:2020orv,Bevilacqua:2021cit} and possible combination of side-band analysis with $t\bar{t}h/t\bar{t}Z$ ratios~\cite{Mangano:2015aow,Goncalves:2021dcu} may change this scenario, controlling the respective uncertainties, and pushing further forward the sensitivity with this extra channel in the near future.

\acknowledgements

We thank Johann Brehmer and Sam Homiller for helpful discussions. RKB and DG thank the U.S.~Department of Energy for the financial support, under grant number DE-SC 0016013. The work of FK is supported by the U.S.~Department of Energy under Grant No.~DE-AC02-76SF00515 and by the Deutsche Forschungsgemeinschaft under Germany’s Excellence Strategy - EXC 2121 Quantum Universe - 390833306. Part of this work was performed at the Aspen Center for Physics, which is supported by National Science Foundation grant PHY-1607611. Some computing for this project was performed at the High Performance Computing Center at Oklahoma State University, supported in part through the National Science Foundation grant OAC-1531128.


\bibliography{references}

\end{document}